\documentstyle[11pt,newpasp,epsf]{article}

\begin{document}

\title{Galaxy Dynamics from Edge-On Late-type Galaxies}
\author{J. J. Dalcanton$^1$ \& R. A. Bernstein$^2$}
\affil{$^1$University of Washington, Box 351580, Seattle WA 98195 \\
$^2$Carnegie Observatories, 813 Santa Barbara St, Pasadena CA 91101}

\begin{abstract}
  We present first results of a program to study the dynamics of
  undisturbed bulgeless, low surface density disk galaxies in order to
  probe the underlying structure of dark matter halos.  High
  resolution H$\alpha$ rotation curves are combined with optical and
  infrared imaging to place strong limits on the halo profiles.  We
  find noticable variation in the shapes of the rotation curves, in
  contrast to previous claims.  The implied density profiles are still
  significantly more shallow than profiles derived from most N-body
  simulations; unlike previous HI observations, beam-smearing cannot
  significantly affect this result.  Based upon stellar mass profiles
  derived from $K^\prime$ band observations, we derive the angular
  momentum distribution of the stellar disk and find it to be broader
  than that of a uniformly rotating solid-body sphere, but remarkably
  consistent from galaxy-to-galaxy.  Finally, based upon $K^\prime$
  band surface brightness profiles, we find that low surface density
  disks must be significantly sub-maximal.  Furthermore, maximal disk fits
  based upon Modified Newtonian Dynamics (MOND) have maximum
  mass-to-light ratios which are too small to be consistent
  with stellar population models; without the ability to significantly
  adjust inclination angles or infrared mass-to-light ratios, this
  sample presents great difficulties for MOND.
\end{abstract}



Although the internal structure of dark matter halos is an extremely
important test of cosmological theories, few secure observational
constraints currently exist.  In disk galaxies, the structure of the
halo is best explored with rotation curves.  However, both the
luminous and dark matter contribute significantly to the enclosed
mass, disguising the dynamics of the dark matter halo, and altering
its structure as well.  Therefore, while rotation curves are the most
sensitive dynamical indicators of a galaxy's total mass distribution,
they are a poor measure of the dark matter profile alone.

More direct probes of the dark matter are provided by low surface
brightness galaxies (LSBs).  There is strong dynamical evidence that
LSBs have low baryonic surface density, and thus the disk contributes
little to the dynamics of the galaxy, and the resulting rotation curve
is dominated by the dark halo; the few LSB rotation
curves published to date rise remarkably slowly, becoming
asymptotically flat only at several disk scale lengths (Goad \&
Roberts 1981, de Blok et al.\ 1996, Makarov et al.\ 1997, van Zee et
al.\ 1997, van der Hulst et al.\ 1993).  Furthermore, LSBs span a wide
range in mass, and thus can be used to trace systematic variations in
the shapes of dark matter halos as a function of mass.


We have been pursuing a study of LSB dynamics using galaxies selected
from the Flat Galaxy Catalog (Karanchetsev et al.\ 1993), a large
sample of edge-on galaxies ($a/b\ge7$, $a\!>\!0.6\arcmin$).  We have
selected 50 galaxies which appear to have low surface brightnesses
when seen edge-on; because these galaxies are optically thin, their
face-on central surface brightnesses will be even lower.  We also
required the galaxies to be completely bulgeless and undisturbed (i.e.
no warps or gross asymmetries).

\begin{figure}
\vbox {
  \begin{minipage}[l]{0.5\textwidth}
    {\centering \leavevmode \epsfxsize=\textwidth 
      \epsfbox{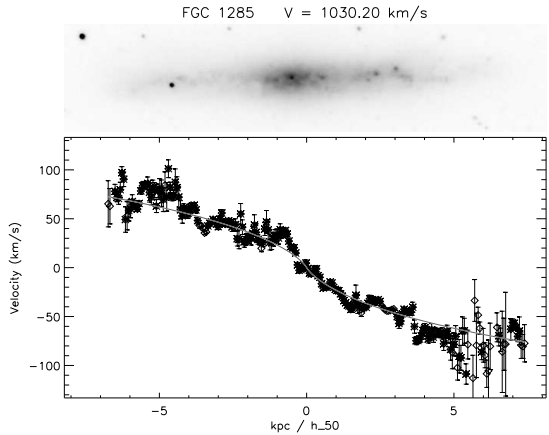}}
  \end{minipage} \ \hfill \
  \begin{minipage}[r]{0.5\textwidth}
    \caption{\scriptsize{$R$ band image (top) and extracted rotation curve 
        (bottom), for a galaxy from the sample.  All plots have been
        scaled to the same horizontal scale.  Open symbols represent
        points of substantially lower signal-to-noise.  In this
        preliminary work, rotation speed is currently extracted as the
        center of the H$\alpha$ line, not at the extrema as would be
        proper for edge-on galaxies.  However, as the line widths are
        $\pm3$ km/s of the instrumental line widths, we expect little
        change when the proper rotation curves are extracted.
        Furthermore, for slowly rising rotation curves, our
        simulations have shown that using line centers as a measure of
        the rotation curve changes the rotation curve by less than
        5\%. }}
  \end{minipage}
}
\end{figure}

We obtained high resolution ($\sim\!1\!-\!1.5^{\prime\prime}$)
H$\alpha$ rotation curves for a subset of 35 of these galaxies.  The
rotation curves accurately probe the dynamics of the galaxies to very
small radii, and at high resolution (0.1-0.5 kpc for the majority of
our sample).  An example image and rotation curve is shown in Figure
1.  We have imaged the sample in $B$, $R$, and $K^\prime$, and
confirmed that all have extremely low surface
brightnesses, in spite of having maximum rotation speeds up to 250
km/s; the median $K$-band surface brightnesses of our sample is more
than 2 magnitudes per square arcsecond fainter than the median of the
de Jong (1995) sample of face on spirals. The majority of these
galaxies are extremely blue ($R-K\!<\!$2.5), lack dust lanes,
and lie on the $B$-band Tully-Fisher relationship for low
luminosity galaxies of Stil (1999), all of which suggests that
extinction is not a significant problem for the majority of the
sample.  Roughly 8 galaxies which have $R-K\!>\!2.5$ have been
eliminated from further analysis, to alieviate any concerns about
extinction affecting the rotation curves.  

The addition of infrared imaging to our sample allows us to accurately
subtract the mass of the stellar disk ($M_*$) from our measured
rotation curves, given the insensitivity of the $K$-band mass-to-light
ratio to variations in star-formation history.  We note, however, that
HI is the largest baryonic contribution to the observed rotation
curve; we find $M_{HI}/M_*\!\sim\!1-4$ for our sample. The rotation
curves are indeed dark matter dominated
($\left<M_{dark}/M_{baryonic}\right>\!\sim\!3.5$ at the last measured
point).

\begin{figure}
\plotone{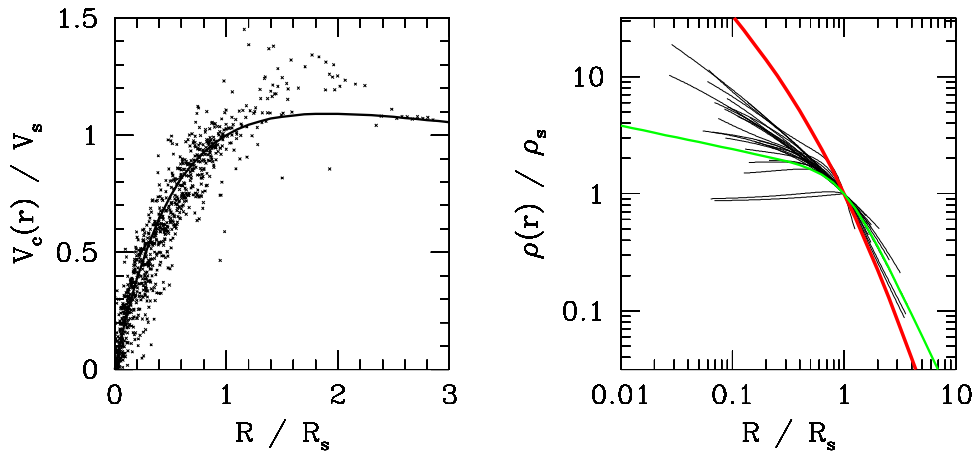}
\caption{\scriptsize{All extracted rotation curves are shown in the
    left panel, scaled by fits to the model suggested by Kravstov et
    al.\ (1998).  The heavy line is the ``universal'' rotation curve
    found by Kravstov et al.  Considerably more scatter is found in
    the high-resolution data than in the HI data plotted by Kravstov
    et al.\ -- 90\% of the data points fall within $\pm$20\% of their
    preferred fit. Some of this scatter is due to the low extinction
    of the galaxies, however (see Figure 1).  The implied density
    profiles are plotted in the right panel.  The heaviest solid line
    is the NFW profile, and the second heaviest line is the shallower
    fit of Kravstov et al.  No baryonic component has been subtracted
    from these plots, and hence the central density profile will
    become shallower.  No galaxies with $R-K\!>\!2.5$ have been
    included, to avoid problems with extinction; 60\% of the galaxies
    remaining have $R-K\!<\!2.0$. Note also that the scaling by $V_s$
    and $R_s$ in the left-hand figure masks the large variation in
    halo density and inner profile shape implied by the density
    profiles in the right-hand figure.}}
\end{figure}


\begin{figure}[t]
\plottwo{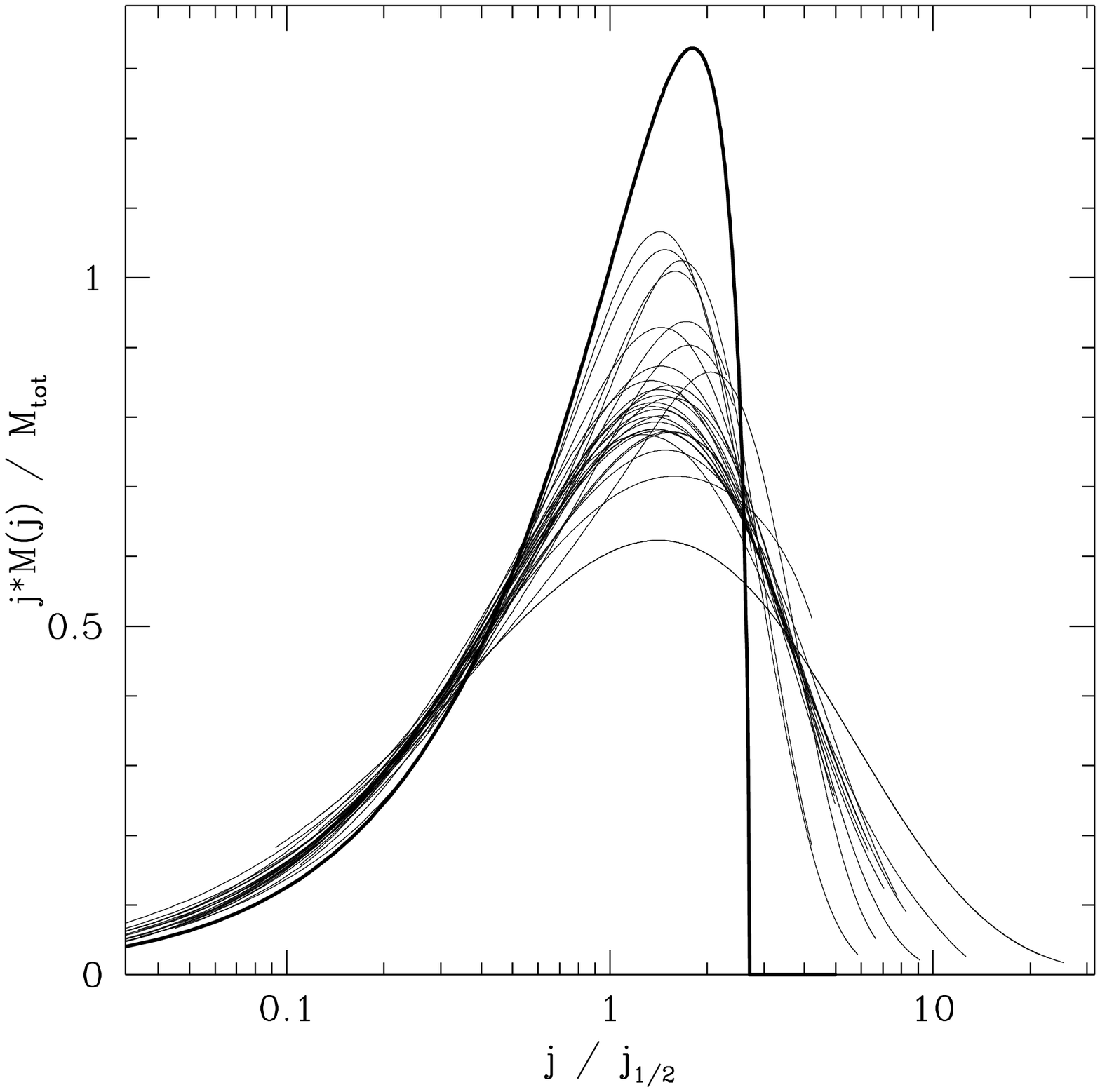}{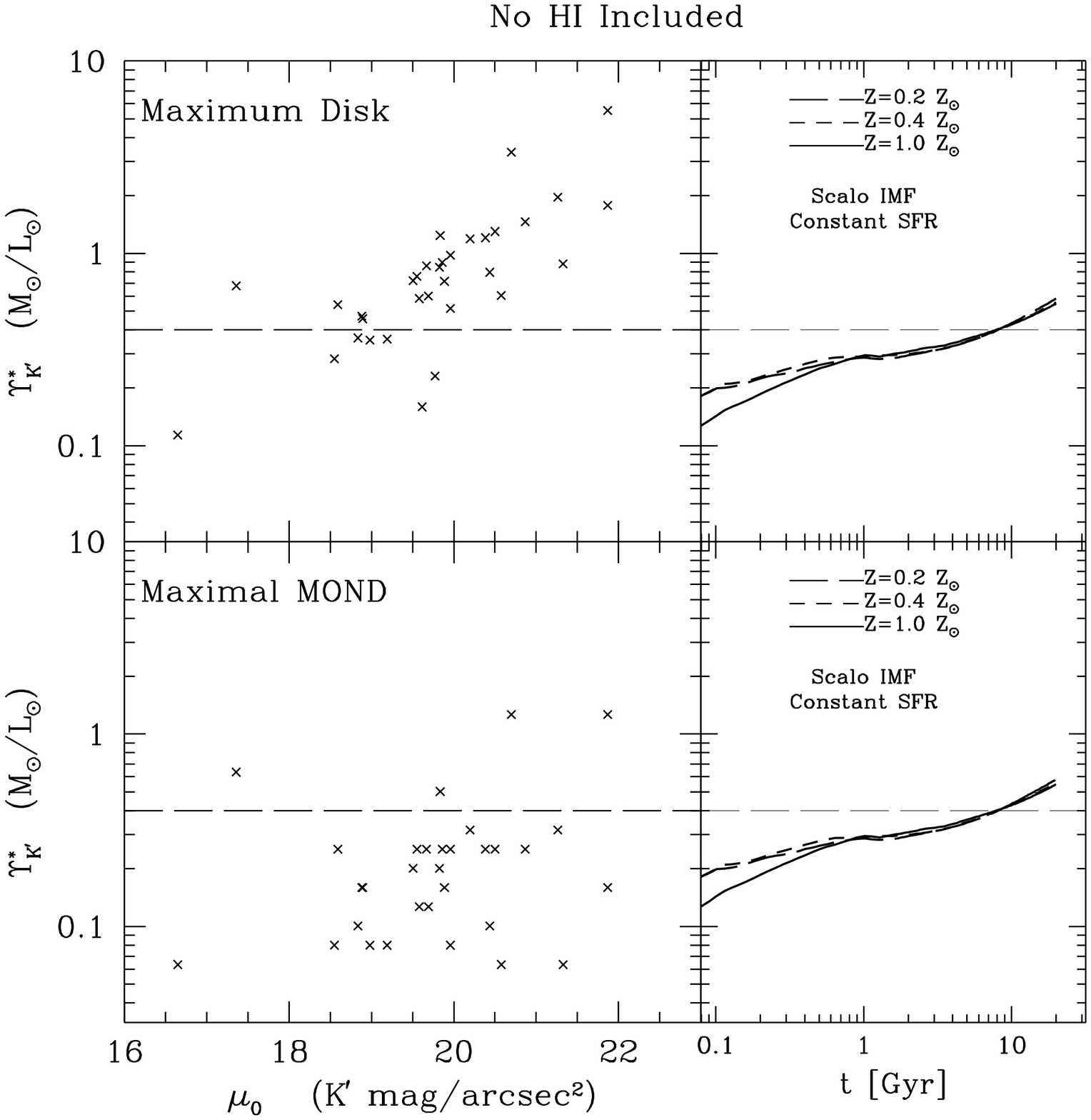}
\caption{\scriptsize{[LEFT] Specific angular momentum
    distributions, based upon exponential disk fits to the $K^\prime$
    images and the fitted rotation curves.  All curves have been
    scaled to $j_{1/2}$, the specific angular momentum containing half
    the mass, and to the same total disk mass.  The heavy curve is the
    distribution expected for a sphere in solid body rotation, as is
    often assumed in disk formation models. HI has not been included.
    [RIGHT] Maximal Mass-to-Light ratios in $K^\prime$ for the stellar
    disk under Newtonian gravity (upper) and MOND (lower).  The right
    panels show expected M/L for a Scalo IMF and constant
    star-formation, for different metallicities (Bruzual \& Charlot
    1999).  Under Newtonian gravity, disks become systematically
    sub-maximal with decreasing surface brightness; the maximal disk
    value of $\Upsilon_{K^\prime}$ is too high to be consistent with
    stellar populations.  Under MOND, the stellar mass-to-light ratios
    are too small to be consistent with stellar population models.
    Including HI will make these limits more severe.}}
\end{figure}

There has been considerable attention paid in the literature to the
apparent contradiction between N-body predictions of steeply rising
rotation curves (cf.\ Navarro et al.\ 1997, Moore 1999; but see
Kravstov et al.\ 1998) and HI observations of slowly rising rotation
curves for dwarf and LSB galaxies (cf. de Blok et al.\ 1997).  While
there seem to be significant conflicts for dwarf galaxies, the
existing data on LSB galaxies was based upon relatively low resolution
HI synthesis observations, and derived dark matter core radii which
were comparable to the resolution of the beam.  Kravstov et al.\ 
(1998) have used the same data to argue that all LSBs have similar
rotation curves and a self-similar halo profile.  In the left panel of
Figure 2 we plot the comparable data for our sample of high-resolution
H$\alpha$ rotation curves.  Note that there is considerable scatter
($\pm$20\%).  The scatter also masks significant variations in the
shape of the rotation curves, as can be seen by the density profiles
derived from the RCs in Figure 2.  Note also that the discrepancy
between LSB observations and the steep cusps predicted by simulations
persists at high-resolution, but that the central rotation curves are
somewhat steeper than the fits of Kravstov et al.\ (1998).

One common feature of models of disk galaxy formation (see Mo, this
volume) is the assumption of detailed angular momentum conservation
for the collapse of a sphere of gas in solid body rotation (Crampin \&
Hoyle 1967).  We test this assumption in the left panel of Figure 3,
where we plot the angular momentum distributions of the stellar disk,
derived from the $K^\prime$ observations and the rotation curve.  The
distributions are remarkably similar, although the data spans a factor
of nearly 5 in rotation speed.  The distributions are also broader
than a sharp-edged sphere, as would be expected for smoother initial
overdensity.  Once the distribution of HI is known, we can calculate
the full baryonic angular momentum distribution.

Finally, we can use the robustness of $K^\prime$ mass-to-light ratios
($\Upsilon_{K^\prime}$) to measure the mass contribution which
baryonic disks make to the overall dynamics.  In the right panel of
Figure 4, we show $\Upsilon_{K^\prime}$ derived from maximum disk fits
to the rotation curve.  The upper panel shows that disks become
progressively ``sub-maximal'' at decreasing mass surface density.  We
have repeated this exercise for MOND dynamics, and find that the
derived values of $\Upsilon_{K^\prime}$ are too low to be consisent
with reasonable star formation histories and normal IMFs.  The
contribution from HI has not been included in these fits, and will
further reduce the allowed values of $\Upsilon_{K^\prime}$, making the
limits for MOND more stringent.

\acknowledgments

We thank the staff of Carnegie Observatories for the generous
allocations of telescope time which have made this project possible,
and Frank van den Bosch and Ben Weiner for interesting discussions.

\end{document}